\documentclass[doublespacing]{elsart}
\usepackage{epsfig}
\usepackage{amssymb}
\begin{document}
\runauthor{dai, yang}
\begin{frontmatter}
\title{The correlations between BL Lacs and ultra-high-energy cosmic rays
deflected by using different GMF models
}
\author[IHEP,utah]{Zhen Cao}
\author[ynu]{Ben Zhong Dai}
\author[ynu,ynau]{Jian Ping Yang}
\author[YAO,ynu]{Li Zhang}

\address[IHEP]{Laboratory for Particle Astrophysics, Institute of High Energy
Physics, Chinese Academy of Sciences, Beijing, 10039, P. R. China}
\address[utah]{Department of Physics, University of Utah, Salt Lake
City, UT 84112, USA}
\address[ynu]{Department of Physics, Yunnan University, Kunming,
650091, P. R. China }
\address[ynau]{Yunnan Agricultural University, Kunming,
650201, P. R. China }
\address[YAO]{National Astronomical Observatories/Yunnan Observatory,
  Chinese Academy of Sciences, P.O. Box 110, Kunming, 650011, P. R. China}

\begin{abstract}
Some studies suggested that a correlation between locations of BL
Lacertae objects (BL Lacs) and the arrival directions of the
ultrahigh energy cosmic rays (UHECRs) exists. Especially by
assuming the primary particles charged $+1$ and using a galactic
magnetic field (GMF) model to calculate the deflections of the
UHECRs, the significance of correlation is improved. We construct
a new GMF model by incorporating all progresses in the GMF
measurements in recent years. Based on a thorough study of the
deflections of the UHECRs measured by the AGASA experiment, we
study the GFM model dependence of the correlation between the
UHECRs and the selected BL Lacs using the new model together with
others. It turns out that only specific one of those GMF models
makes the correlation significant, even if neither GMF models
themselves nor deflections of the UHECRs are not significantly
different. It indicates that the significance of the correlation,
calculated using a method suggested in those studies, is
intensively depending on the GMF model. Great improvement in
statistics may help to suppress the sensitivity to the GMF models.

\end{abstract}

\end{frontmatter}

\section{Introduction}
Search for sources of ultrahigh energy cosmic rays (UHECRs) is
very important and many  researches have been done recently,
including searches for correlations between the UHECRs measured by
AGASA and HiRes experiments and BL Lacertae objects (BL Lacs)
\cite{BL1,BL:GMF,BL2}. Identifying the sources will provide us
direct information  about the acceleration mechanisms of the
UHECRs that is essential to understand phenomena such as
Greisen-Zatsepin-Kuzmin (GZK) ¡®cutoff¡¯\cite{Greisen,Zatsepin}.
The distribution of arrival directions of UHECRs, however, is
remarkably isotropic. There is no correlation between astronomical
objects and cosmic rays being confirmed. Arrival directions should
not point back to sources unless the UHECRs are neutral, e.g.
$\gamma$'s or neutrinos. The charged cosmic ray primaries will be
deflected while propagate through the galactic and intergalactic
magnetic fields (GMF and IGMF). UHECRs above $4\times10^{19}$~eV
might be considered  for small-scale anisotropy searching.
Deflections of them would be so small that the uncertainty of the
deflection estimation would not be harmful. If there was any
correlations between the UHECRs and any known objects, the
correlation should not be strongly dependent on specific GMF
models. However, this has  yet to be seriously tested. This paper
is devoted  to test the model dependence. At first, a new GMF
model is constructed by incorporating all recent progresses in GMF
measurements. Secondly, the deflections of UHECRs using different
GMF models, including the models used for the AGASA data analysis,
are compared. Finally, the correlations between the UHECRs and the
selected BL Lacs are investigated to check the influences from the
different GMF models.

 Active Galactic Nuclei(AGNs) have been considered as
UHECR sources by
authors~\cite{Berezinsky90,Rachen,Berezinsky:2002nc}. The AGASA
data set \cite{takeda} has exhibited clustering in the
experimental resolution. Some authors have suggested that the
clusters may be due to point sources (although no cluster has been
confirmed by the HiRes experiment\cite{Abbasi1,Abbasi2} ). It has
been argued that those clusters might be aligned with BL Lacs
\cite{BL1,BL:GMF,BL2}. In those papers, significant correlations
are expressed no matter what primary particles of those UHECRs are
assumed, neutral or charged. In case of the proton primary is
assumed, deflections of the protons in the GMF are estimated using
a specific model. Even more significant correlation is claimed
with the proton primary assumption.

BL Lacs are blazars (AGNs with relativistic jets directed along
the line of sight) characterized in particular by absence of
emission lines.  This indicates low ambient matter, therefore it
is a favorable condition for accelerating particles to ultra high
energies. Studies have suggested that the acceleration of
particles in jets can be explained using pinch mechanism and the
maximum energy of particles can be greater than $10^{20
}$eV\cite{Veniamin}. The AGASA results seem to enhance the
hypothesis. However, the correlation between the UHECRs and the BL
Lacs must be concretely confirmed. If the UHECRs are not neutral,
the GMF model dependence needs to be investigated thoroughly
before drawing any conclusion on this mechanism.

Deflections  of charged UHECRs in the intergalactic magnetic
fields are assumed random and  unpredictable due to lack of
knowledge about the magnetic fields. The deflections  in the GMF
are better understood because the knowledge about the GMF  is
greatly enhanced in the last ten years  using rotation measures
(RMs) of radio polarization from pulsars and extragalactic radio
sources. The RM data reveals many new features of the GMF, such as
a central rotating bar in the galactic plane, a dynamo structure
in the galactic halo, a magnetic dipole at the galactic center
(G.C.) and so on. The GMF model could be improved by taking all of
the features into account. There are many GMF models available in
the market depending on implementations of those observational
facts. This offers an opportunity for investigating the model
dependence of the correlation between the BL Lacs and the UHECRs.
The GMF model is improved by incorporating the latest updates of
the RM measurements in this paper, then the GMF model dependence
is studied using the AGASA data and the new model and other
available GMF models. This study is essential for further
investigation using HiRes data.

This paper is organized as following.  Improvements of the GMF
model and differences between models are described in
Sect.~\ref{sect:GMF}.  Deflections of the UHECRs using different
GMF models are analyzed in Sect.~\ref{sect:DF}. Correlations
between the BL Lacs and the UHECRs are re-estimated for the GMF
model dependence testing in Sect.\ref{sect:CA}. Conclusions are
drawn based on the comparisons in Sect.~\ref{sect:conclusions}.

\typeout{SET RUN AUTHOR to \@runauthor}

\section{Galactic Magnetic Field}
\label{sect:GMF}
The Galactic Magnetic Field (GMF) is composed of regular field and
turbulent field. The regular field keeps no change in time and
distributes close to matters in our galaxy. The turbulent field
that is due to localized activities of objects in our galaxy.
Recent studies \cite{han04paper} has discovered that the turbulent
fields exist on all scales from a few pc's to the whole galaxy.
Strengths of the turbulent fields can be up to twice of the
regular field at the same place. Directions of the turbulent
fields are isotropic. To model the turbulent fields, their
directions are randomly chosen and  their strengths are randomly
sampled from a half to twice of the regular field strength at the
same place.

The regular field includes three components according to their
sources and regions where they are distributed. The main component
is located inside the disk of our galaxy where a majority  of the
galactic matter is distributed in a spiral structure. The magnetic
field is also distributed in the same 2-dimensional spiral
structure. Outside the galactic plane, the filed strengths
decrease with the distance from the plane as the galactic matter
distribution does. Directions of the fields are parallel to the
disk. In the galactic halo,  a pair of toric structures of the GMF
are found. The toruses are located  in regions about 4kpc from the
disk in both upper(north) and lower(south) hemispheres. Directions
of the toric fields are also parallel to the disk, rotating in a
clockwise direction in the upper hemisphere but reversed in the
lower. Moreover, evidences indicate that there is a dipole field
at the center of our galaxy. All three components are described in
details in the following subsections.

  \subsection{ The 2-dimensional magnetic field component in the galactic disk}
\label{mf} The field is  distributed in a spiral structure as the
baryonic matter does in the area beyond 4 kpc from the G.C. The
model of the magnetic field component in the galactic disk is
described by the following parameters :
   \begin{itemize}
\item Distance from the Sun to the Galactic center, $R = 8.5$ kpc,
\item Local  field strength, $B_0 = 1.4 \,\mu {\rm
G}$\cite{han04paper,Beck00}, \item Pitch angle $p= -
8.2^\circ$\cite{Beck00,Han94,Han01}, \item Distance from the sun
to the first field reversal $d$=-0.2kpc \cite{Han94,Han01},.
\end{itemize}

The field strength at a point ($r,\theta$) in the galactic disk is

\begin{equation}
B(r,\theta) = B(r)\, \, \cos\left(\theta - \beta
\ln\left(\frac{r}{R}\right) + \phi \right) \,\,\,\,\,(r> 4kpc),
\label{Bd1}
\end{equation}
where $\beta \equiv 1/\tan(p)$, the constant phase $\phi$ is given by
\begin{equation}
\phi = \beta \ln\left(1 + \frac{d}{R}\right) - \frac{\pi}{2} \, ,
\end{equation}
and $B(r)$ describes the change of the strength with distance from
the center of our galaxy. The strengths of the fields
exponentially fall off with the distance from the G.C. and  reads
as
\begin{equation}
B(r) = \frac{B_0\:exp(- \frac{(r-R)}{r_b})}{\cos (\phi )}
\,\,\,\,\,(r>4kpc), \label{Bd3}
\end{equation}
where the scale radius $ r_b=7 $ kpc \cite{han-lecture}. The
direction of the field at $(r,\theta)$ is determined by the pitch
angle and described by the two components
\begin{equation}
B_\theta = B(r,\theta)\, \cos(p), \qquad B_r = B(r,\theta)\,
\sin(p).
\end{equation}


Recently, there is a clear evidence from GLIMPSE (Galactic Legacy
Mid-Plane Survey Extraordinaire) about a bar-like structure,
consisting of relatively old and red stars, in the central area of
our galaxy. The bar is about 27,000 light years ($\sim$8kpc) long,
longer than previously believed. This survey also shows that the
bar is oriented at 45$^\circ$ respect to a line connecting the sun
and the center of the galaxy\cite{Benjamin}. This bar-shaped
structure has been built in the new GMF model for $r<4kpc$. A
Gaussian is used for the description, i.e.
\begin{equation}
B(r,\theta) = 4 \,\cdot exp(
-\frac{r^2}{4}(sin\theta-cos\theta)^2) \,\,\,\,\,(r< 4kpc),
\end{equation}
where the field strength is 4 $\mu$G in the bar area that is about
1kpc wide.

\subsection{ The magnetic fields outside the galactic disk}

There are two components in the region of $z\ne 0$. The first part
is the extension of the spiral and bar-shaped field. Outside the
galactic disk, the strengths of the fields decreases following a
Gaussian function of $z$, i.e.
\begin{equation}
B(r,\theta,z) = B(r,\theta)\,exp( -\frac{1}{2}(\frac{z}{h})^2),
\end{equation}
Where the Gaussian width $h$=0.6 kpc\cite{Han-priv} being
 consistent with the matter distribution. Directions of the fields
 are the same as the 2-dimensional fields at $z=0$ in both
 the upper and lower hemispheres. This is a natural description.

The magnetic fields in the halo near the galactic disk are
dominated by the extensions of the spiral field. In regions
further away from the disk, the magnetic fields are found to have
a dynamo-like structure which is described better by a toric field
model\cite{Han01}. The magnetic fields are distributed in two
toruses located in the upper and lower hemispheres, respectively.
The fields of the two toruses are in opposite directions. The
centers of the toruses are about $\pm$1.22 kpc from the disk and
are parallel to the disk. There are also evidences showing that
the strengths of the fields are about 1$\mu {\rm G}$ around z=1.5
kpc\cite{Han99}. The toric fields can be modelled by
\begin{equation}
B(r,\theta,z)=B_t exp(
-\frac{1}{2}(\frac{r}{r_t})^2)[exp(-\frac{1}{2}(\frac{z-1.22}{h_t})^2)-exp(-\frac{1}{2}(\frac{z+1.22}{h_t})^2)]
\end{equation}
where $r_t$ is 8.5 kpc, $h_t$ is 1.22 kpc and $B_t$ is 1.85 $\mu$
G.

  \subsection { dipole magnetic fields located at  the galactic center}

 The local magnetic fields in the vicinity of the
solar system have been found to have a small vertical component
about 0.3 $\mu {\rm G}$ and pointing from the South to the North
\cite{Han94}. This observational fact can be interpreted as that
there exists a dipole at the center of our galaxy and points to
the north. The dipole fields are
\begin{equation}
\label{r31}
 B_x = -3 \mu {_{\rm G}} \sin(\theta) \cos(\theta) \cos(\phi)/r^3
\end{equation}
\begin{equation}
\label{r32}
 B_y = -3 \mu {_{\rm G}} \sin(\theta) \cos(\theta) \sin(\phi)/r^3
\end{equation}
\begin{equation}
\label{r33}
 B_z =  \mu {_{\rm G}} (1-3\cos^2(\theta))/r^3
\end{equation}
where the constant $\mu {_{\rm G}}$ must be  184.2 $\mu {\rm G}
(kpc)^3$, and $r$, $\theta$, $\phi$ are spherical coordinates at
the center of galaxy and $\phi=0$ is the direction of the sun.
This is a  very strong dipole moment  determined only by a small
value measured at 8.5 kpc away from the dipole. It has to be
tested with great care.

There is not enough experimental data to provide sufficient
constraint on the model beyond 20kpc from the G.C., therefore the
new model is only valid in a range of 20 kpc from the center of
our galaxy in all directions.

In order to compare the new model with others, two models that are
used in Ref. \cite{BL:GMF} (model-T) and Ref.
\cite{HAJIME,Prouza,Jaime,Hiroyuki} (model-M) are chosen. Those
models are widely used in the correlation analyses between the BL
Lacs and the UHECRs.

\begin{figure}
\begin{center}
\includegraphics[width=9cm]{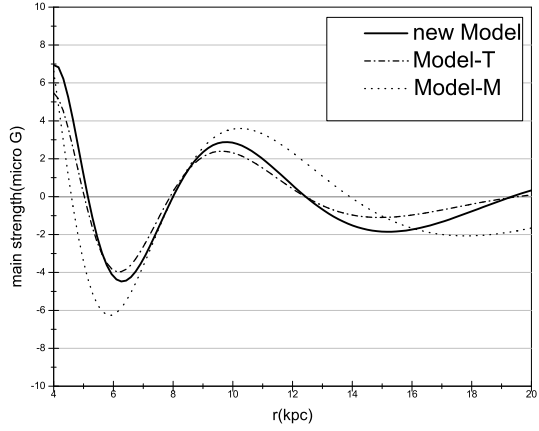}
\caption{ The change of the galactic magnetic field strength as a
function of the distance from the G.C., r, in the direction of the
 sun, $\phi=0$. }
 \label{GMF_r}
\end{center}
\end{figure}

At first, the fields in the galactic disk are compared. In
Fig.~\ref{GMF_r}, the field strength is plotted against the
distance from the G.C. along the direction $\phi=0$. The model-T
has very similar behavior compared with the new model beyond 4
kpc, including the strengths, reversals and the phases of the
fields. This similarity indicates that the spiral structure and
the reversals of the fields are essentially same in those two
models. The model-M shows a slightly different behavior. The
reversals occur at different locations and more importantly the
field strengths remain quite strong at large distances, e.g. near
20kpc from the G.C., where the field strengths are reduced to very
weak in the other two models. The rather large and flat field
strengths between 16 kpc and 20 kpc may have strong consequences
in deflections of the UHECRs. The other major difference is that
the latest progress about the bar-shaped baryonic matter
distribution near the G.C. has been incorporated in the new model.
It may cause significant differences in the deflections of UHECRs
passing by the G.C.

\begin{figure}
\begin{center}
\includegraphics[width=9cm]{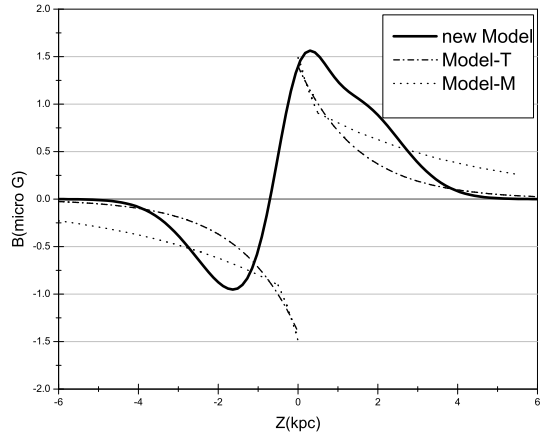}
\caption{ The change of the galactic magnetic field strength as a
function of the distance from the galactic disk, z, in the
vicinity of the sun.}
 \label{GMF_z}
\end{center}
\end{figure}

It is noticed  that the strengths have discontinuities at $z=0$ in
the descriptions of the halo fields in both the model-T and
model-M. Those discontinuities occur also at all reversals of the
fields in the galactic disk. Those discontinuities are not natural
and cause the deflections of the UHECRs in different ranges to
cancel each other. In the new model the strength has a smooth
transition cross the disk at $z=0$ everywhere in our galaxy. In
Fig.~\ref{GMF_z}, the GMF strengths are plotted as functions of
distances from the disk in the vicinity of the sun. It is
important to note that the halo fields have similar behaviors
beyond 4 kpc from the galactic disk in the new model and the
model-T, however, the model-M assumes that the fields are quite
strong in a larger region of the halo. The strengths of the fields
are much stronger than other models beyond 3kpc from the disk.
Those features cause much stronger deflections for most UHECRs
that come from high galactic latitudes.


\section{Analysis of deflections using three GMF models}
\label{sect:DF} A deflection angle  is defined as the difference
between the observed UHECR  direction and the primary direction of
UHECR i.e. the direction outside our galaxy which is calculated by
using time reversing symmetry by assigning a charge to the cosmic
ray particle, e.g. assuming it is a proton. The 57 UHECR events
observed by AGASA experiment ($E>4 \times 10 ^{19} eV$) adopted
from  Ref. \cite{takeda} are used as examples for studying the
deflections of the UHECRs in the GMF and model dependence in this
paper.

\begin{figure}
\begin{center}
\includegraphics[width=9cm]{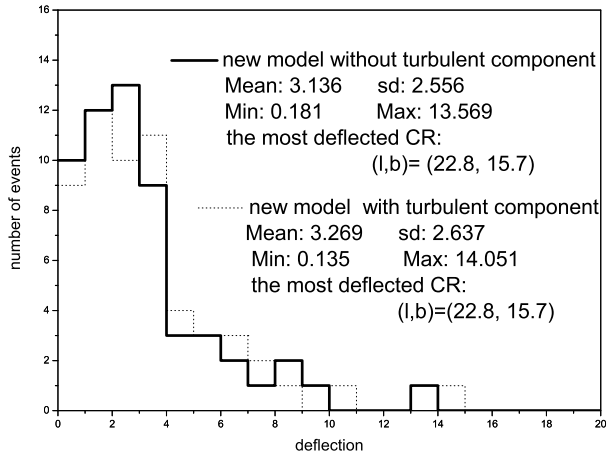}
\caption{ The distribution of deflection angles of 57 AGASA events
using the new GMF model. The thick solid histogram represent the
case that the turbulent component is turned off, while the thin
histogram represent that the turbulent component of the GMF is
considered. }
 \label{our_defl}
\end{center}
\end{figure}

Using the new GMF model, most of  AGASA events are found to be
bent less than $10^\circ$ with a peak at  $3^\circ$, only one
event is bent about $13^\circ$( Fig.~\ref{our_defl} ). With or
without the  turbulent components of the GMF, the deflection
angles of the cosmic ray samples  are compared in the same figure.
It is shown that turbulent fields systematically shift the
deflection angle by about $0.1^\circ$. It is so small compared
with the effect of different GMF models that  the turbulent
component is negligible for the UHECR samples. The average bending
angle is 3.16$^\circ$.

\begin{figure}
\begin{center}
\includegraphics[width=9cm]{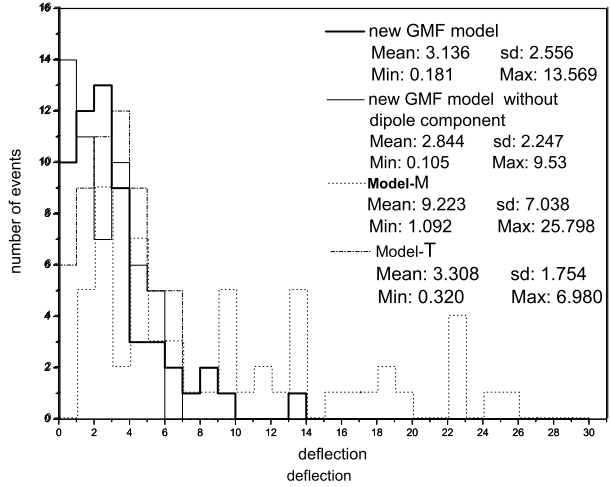}
\caption{ The distribution of deflection angles of the 57 AGASA
events different GMF models. Thick solid line is for the new
model, thin solid line is for the new model without the dipole
component, dotted dash line is for the model-T and dot line is for
the model-M. }
 \label{four_defl}
\end{center}
\end{figure}

The dipole field at the galactic center is so strong that any
particles passing by the G.C. could be bent severely. In
Fig.~\ref{four_defl}, the deflections are compared with or without
the dipole fields. The most deflected cosmic ray is from the
direction with the galactic longitude and latitude of
(l,b)=(22.8$^\circ$,15.7$^\circ$) and is bent by 13.6$^\circ$
using the new model. The bending is mainly caused by the dipole
field, since the trajectory is close to the G.C., i.e. the angle
between the trajectory and a connection from the G.C. to the sun
is about 27$^\circ$. Once the dipole field is turned off in the
deflection calculation, this event is bent with a smaller angle,
while the most deflected cosmic ray is replaced by another from
(l,b)=(154.5$^\circ$,15.6$^\circ$) and is bent by 9.53$^\circ$.
According to this analysis, one should cautiously treat those
events that their trajectories are close to the G.C. with models
that the dipole fields are included. To avoid a great uncertainty
associated with the dipole component, those events that pass by
the G.C. with close distances should be cut.

For comparison, the model-T and model-M are used to the same data
set for deflection estimation. Deflections of all events are
confined to $7^\circ$ using model-T (dotted dash line in
Fig.~\ref{four_defl}). The average bending angle is about
3.3$^\circ$ which is not significantly different from  the new
model. Within the statistic fluctuations, one might draw a
conclusion that the deflection estimations are essentially same by
using those two models. However, using the model-T, the bending
angles are populated around about $3^\circ\sim 4^\circ$, which
differ from the new model. Using the new model, most of cosmic
rays are bent with angles less than 4$^\circ$. 5 cosmic ray events
are bent  more than 7$^\circ$, which are caused by the dipole
component. There is no event being bent more than 7$^\circ$ if the
dipole component is turned off in the new model.

 Deflections are scattered in a
wide range using model-M. Many cosmic rays are bent more than
$10^\circ$ (dotted line in Fig.~\ref{four_defl}). Statistically,
this model has different behavior from the new model and the
model-T. The wide distribution is mainly caused by the halo field.
From Fig.~\ref{GMF_z}, one can see that the halo field of model-M
has a very large tail extending 8kpc away from the disk. On the
other hand, both new model and the model-T have similar behaviors
beyond 4kpc from the disk, so that the comic rays have the similar
deflections even if the halo fields  are very different near the
disk. This indicates that the fields in the halo, for instance
beyond 4kpc, are crucial for UHECR deflections. It is expected
that the GMF measurements in the high galactic latitude region
will be improved in the future. They are essential for
understanding the cosmic ray deflections.

\section{The  correlation  between UHECRs and BL Lacs}
\label{sect:CA} Using the chance probability function $p(\delta)$
introduced in Refs.~\cite{BL1,BL:GMF,BL2,BL3,BL4}, the
correlations between the UHECR events and sources were
quantitatively estimated for given angular interval $\delta$. The
probability is calculated by counting how often the numbers of MC
events and source pairs are equal or greater than numbers of real
events and source pairs, i.e. $p(\delta)=\frac{ number\; of \;
trial \; sets\; with \;N_{mc} \geq N_{real}}{total \;number\; of\;
trial \;sets\; }$, where the $N_{real}$ and the $N_{mc}$ are
numbers of pairs of real cosmic rays matching with sources and
simulated events matching with sources, respectively. A pair is
defined as an event falls in a circle with an angular radius of
$\delta$ centered any source in a selected samples. The smaller is
this probability, the more significant the correlation is. The
Monte-Carlo events are generated in the horizon reference frame
with a geometrical acceptance $dn \propto \cos
\theta_z\sin\theta_z d\theta_z$, where $\theta_z$ is zenith angle
and the coordinates are transformed into equatorial frame assuming
random arrival time. All events are generated with zenith angles
$\theta_z < 45^\circ$, same as the real event samples are
selected\cite{Uchihori}.  The deflections of both data and MC
samples are calculated by assuming pure proton primary. If the
function $p(\delta)$ exists a minimum  nearby the experimental
resolution(about $2.5^\circ$ for the AGASA experiment), it
indicates a correlation between the observed UHECR's  and the
sources. The value of the $p(\delta)$ is an estimation of the
chance probability. Ten thousand sets of the MC events are
generated for each case.

\begin{figure}
\begin{center}
\includegraphics[width=9cm]{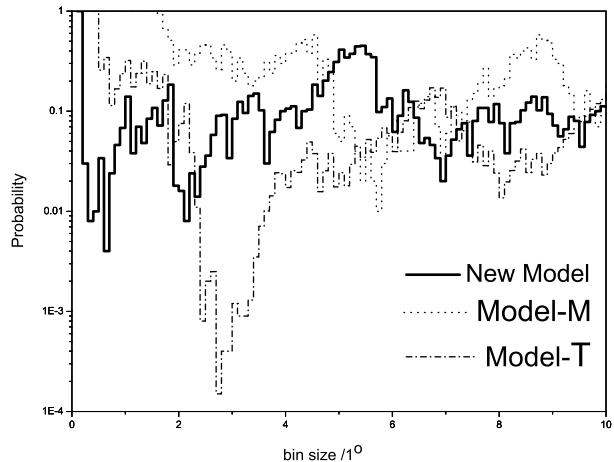}
\caption{ $P(\delta)$ for the set of 178 BL Lacs  (${\rm mag} <
18)$ and AGASA 57 events ($E>4\times10^{19}$~eV). The new GMF
model (solid), model-T (dash-dotted) and model-M (dotted) are used
for the cosmic ray deflection calculation.
 }
 \label{our_gmf}
\end{center}
\end{figure}

From the data of  QSO catalog \cite{veron}, 178 BL Lacs are
selected according to a criterion of   apparent magnitude less
than $ 18$ Ref.\cite{BL:GMF}.  The focus of this paper is the GMF
model dependence of the correlation between selected BL Lacs and
the 57 AGASA events with deflections. First of all, the result in
Ref.\cite{BL:GMF} is reproduced using the model-T as the
dash-dotted line in Fig.~\ref{our_gmf}. It clearly shows a minimum
around $\delta =2.7^{\circ}$. Based on this, the authors claimed a
significant correlation between the UHECRs and the selected BL Lac
samples. However, the same chance probability calculated using the
new GMF model and model-M do not show any correlation between the
AGASA cosmic ray events and the BL Lac samples as solid and dotted
lines in the same figure. This shows that the GMF model dependence
is nontrivial.

As described in Sec.\ref{sect:GMF}, the models themselves are very
similar between the new model and the model-T, except for the
fields both in the galactic disk and halo is improved based on
more modern observations in the new model. The artificial
discontinuities at $z=0$ are replaced by a smooth transition
between the upper and lower hemispheres and the singularity at
$r=0$ is overcome by introducing the bar-structure in the central
area in the new model.

Those small improvements are not expected to significantly change
the deflections of UHECRs. This has been shown in the comparison
in Sec.\ref{sect:DF}. Although the deflections are not exactly
same, within statistical fluctuation, the deflection angle
distributions from those two models agree well with each other.
This also shows that the deflection of UHECRs are correctly
treated which maintains the consistency.

It is noted that the GMF model-T does not include the dipole field
component. It is also aware of that the dipole field component
might be determined with a great deal of uncertainty. Therefore,
the deflection behaviors are studied using the new model without
the dipole component. From the Fig.\ref{four_defl}, it is found
that the distribution of the deflection angles is even more close
to that from the model-T. Only for those the deflection angles are
less than 1$^\circ$, the two models have different behaviors.
Without the dipole component, a similar analysis of the
correlation between the UHECRs and the BL Lacs is still not
significant.

To understand the discrepancy between  results of the correlation
analyses from different models, the numbers of the event-source
matching pairs around the AGASA resolution $2.5^\circ$ are listed
for different GMF models in Table \ref{tab:probability}. The
differences between the new model and the model-T are small.
However, the chance probability jumps  2 orders of magnitudes from
10$^{-4}$ to 10$^{-2}$. This indicates that the function
$p(\delta)$ is too sensitive. A usual problem of such a sensitive
method is poor stability or robustness. It may be very useful for
a big sample of UHECRs where more matching pairs are expected.

It is noticed that the numbers of pairs are not  affected by the
dipole field component.

\section{Conclusions}
\label{sect:conclusions}

Using AGASA 57 UHECR ($E>4 \times 10 ^{19} eV$) and the new GMF
model, model-T and model-M, the GMF model dependence of the
correlation analysis between the BL Lacs and the UHECRs are tested
in this paper. To calculate the deflection of the UHECRs , charge
$Q=+1$ is assigned to all events and deflections by IGMF are
assumed negligible. Using the model-M, the deflections are found
significantly larger than other two models. The reason is that the
magnetic fields in the galactic halo extend much further from the
galactic disk than the other models. The deflection behaviors of
the UHECRs are very similar using the new GMF model and the
model-T. Within the statistical fluctuation, the two models agree
with each other in terms of the deflection angle distributions.
However, the numbers of event-BL Lac matching pairs by using those
two models are different, i.e. slightly larger than one Poissonian
standard deviation. This difference causes a discrepancy of about
2 orders of magnitudes in the chance probability function between
the two models. Both the new GMF model and the model-M does not
support correlations between the AGASA UHECRs and the selected BL
Lacs. This indicates that the correlation analysis method might be
too sensitive to the model, especially at the stage where there is
not enough statistics for the UHECR samples. To draw a conclusion
on the correlation between the UHECRs and the BL Lacs, there needs
to be many more UHECR event samples. There have been lots of UHECR
events collected by the HiRes Experiment and Auger Experiment
recently. To complete this study, however, more constraints on the
GMF models based on further observations are crucial according to
the discussion in this paper.

\begin{table*}
\caption{The numbers of UHECR-BL Lac matched pairs around
$2.5^\circ$, which is the angular resolution of the AGASA
Experiment.}
 \label{tab:probability}
\renewcommand{\arraystretch}{1.5}
\centering
\begin{tabular}{|l|c|c|c|l}
 \hline \hline
angle & new model & model-T  & model-M&
\\ \hline
\footnotesize $2.3^\circ$    &\footnotesize 12 &\footnotesize  12 %
&\footnotesize  6  \\
\footnotesize $2.4^\circ$    &\footnotesize 12 &\footnotesize  15 %
&\footnotesize  6 \\
\footnotesize $2.5^\circ$    &\footnotesize 12 &\footnotesize  15 %
&\footnotesize  8   \\
\footnotesize $2.6^\circ$    &\footnotesize 12 &\footnotesize  16 %
&\footnotesize  9  \\
\footnotesize $2.7^\circ$    &\footnotesize 12 &\footnotesize  18 %
&\footnotesize  9  \\
\footnotesize $2.8^\circ$    &\footnotesize 12 &\footnotesize  19 %
&\footnotesize  11  \\
 \hline \hline
\end{tabular}
\end{table*}

\section*{Acknowledgments}
{\tolerance=400 We are grateful to Dr. J. L.Han for his valuable
comments and discussions with us. This work is supported by NSFC
under contract 10445001, Knowledge Innovation fund (U-526) of
IHEP, China and by Hundred Talents \& Outstanding Young Scientists
Abroad Program of CAS, China(U-610). }

\end{document}